# Explainable Text Classification Techniques in Legal Document Review: Locating Rationales without Using Human Annotated Training Text Snippets

Christian Mahoney, Esq.
e-Discovery
Cleary Gottlieb LLP
Washington DC, USA
cmahoney@cgsh.com

Peter Gronvall
Data & Technology
Ankura Consulting Group, LLC
Washington DC, USA
peter.gronvall@ankura.com

Nathaniel Huber-Fliflet
Data & Technology
Ankura Consulting Group, LLC
Washington DC, USA
nathaniel.huber-fliflet@ankura.com

Jianping Zhang
Data & Technology
Ankura Consulting Group, LLC
Washington DC, USA
jianping.zhang@ankura.com

*Abstract*—US corporations regularly spend millions of dollars reviewing electronically-stored documents in legal matters. Recently, attorneys apply text classification to efficiently cull massive volumes of data to identify responsive documents for use in these matters. While text classification is regularly used to reduce the discovery costs of legal matters, it also faces a perception challenge: amongst lawyers, this technology is sometimes looked upon as a "black box." Put simply, no extra information is provided for attorneys to understand why documents are classified as responsive. In recent years, explainable machine learning has emerged as an active research area. In an explainable machine learning system, predictions or decisions made by a machine learning model are human understandable. In legal 'document review' scenarios, a document is responsive, because one or more of its small text snippets are deemed responsive. In these scenarios, if these responsive snippets can be located, then attorneys could easily evaluate the model's document classification decisions – this is especially important in the field of responsible AI. Our prior research identified that predictive models created using annotated training text snippets improved the precision of a model when compared to a model created using all of a set of documents' text as training. While interesting, manually annotating training text snippets is not generally practical during a legal document review. However, small increases in precision can drastically decrease the cost of large document reviews. Automating the identification of training text snippets without human review could then make the application of training text snippet-based models a practical approach. This paper proposes two simple machine learning methods to locate responsive text snippets within responsive documents without using human annotated training text snippets. The two methods were evaluated and compared with a document classification method using three datasets from actual legal matters. The results show that the two proposed methods outperform the document-level training classification method in identifying responsive text snippets in responsive documents. Additionally, the results suggest that we can automate the successful identification of training text snippets to improve the precision of our predictive models in legal document review and thereby help reduce the overall cost of review.

## I. INTRODUCTION

In modern litigation, attorneys often face an overwhelming number of documents that must be reviewed and produced over the course of a legal matter. The costs involved in manually reviewing these documents has grown dramatically as more and more information is stored electronically. As a result, the document review process can require an extraordinary dedication of resources: companies routinely spend millions of dollars sifting through and producing responsive electronically stored documents in legal matters [1].

For more than ten years, attorneys have been using machine learning techniques like text classification to efficiently cull massive volumes of data to identify responsive information – this is information that is responsive to a party's document requests or requests for production. In the legal domain, text classification is typically referred to as predictive coding. Attorneys typically employ predictive coding to identify so-called "responsive" documents, which are materials that fall within the scope of some 'compulsory process' request, albeit a discovery request, a subpoena, or an internal investigation. While predictive coding is regularly used to reduce the discovery costs of legal matters, it also faces a perception challenge: amongst lawyers, this technology is sometimes looked upon as a "black box." Put simply, typically no extra information is provided for attorneys to understand why documents are classified as responsive.

The research in this study addresses a still not widely studied component of the predictive coding process: <u>explaining why documents are classified as responsive</u>. Attorneys typically want to know why certain documents were determined to be responsive by a predictive model, but sometimes those answers are not obvious or easily divined. This can be confusing for an attorney, especially if the text content of the document does not appear to contain obvious responsive content.

The focus of the research in this study was to develop better performing modeling methods that target the document text that the predictive model identified and used to make the classification decision. In addition, our research in [6] demonstrated that predictive models created using annotated training text snippets perform better than models created using the entire text of the training documents. The precision of snippet-trained models was higher than that of models created with the full documents' text. Despite this performance increase, it wasn't practical to ask humans reviewers to annotate the text snippet to train the model since this task takes more time than assessing the whole document and labeling it as responsive or not – time typically drives costs in legal document reviews. Given this time/cost versus performance improvement dilemma, another focus of our research was to find a way to automate finding text snippets for training without using human reviewers.

The Artificial Intelligence (AI) community has been researching explainable Artificial Intelligence since the 1970s, e.g. medical expert system MYCIN [2]. More recently, DARPA proposed a new research direction for furthering research into Explainable AI (XAI) [3]. In XAI systems, actions or decisions are human understandable – "machines understand the context and environment in which they operate, and over time build underlying explanatory models that allow them to characterize real world phenomena." Similarly, in an explainable machine learning system, predictions or classifications generated from a predictive model or predictive models are explainable and human understandable.

Understanding a model's classification decision is challenging in text classification because of the factors a model considers during the decision-making process, including word volume within text-based documents, and the volume and diversity of words established during the text classification process. In the legal domain, where documents can range from one-page emails to spreadsheets that are thousands of pages long, the complexity of the models creates challenges for attorneys to pinpoint where the classification decision was made within a document.

In legal document review, a document is considered responsive when any portion of a document contains responsive information. This is not always true for many other text classification tasks. For example, in topic classification, when a document is classified to a topic, the entire document may talk about the topic. For the purposes of this paper, we considered any portion of a document (throughout the rest of this paper referred to as a: text snippet) to be a small passage of words within a document.

This paper proposes two machine learning methods for locating responsive snippets within a responsive document. In legal document review, labeling responsive snippets is a high cost and tedious task and not a reasonable request for reviewers. Therefore, in legal document review engagements, we typically have labeled documents, but no labeled responsive text snippets for training. The two proposed methods do not use human-labeled snippets to train models for responsive snippet detection. We conducted experiments, using three datasets from actual legal matters, to evaluate the performance of the two proposed methods and compared their results with the performance of a document-level training classification method. Our research demonstrated that the two proposed methods are effective and outperform the document-level training classification method in identifying responsive text snippets (rationales) in responsive documents. Our main contributions are i) we proposed two innovative methods for training models that identify responsive text snippets (rationales) within responsive documents without using annotated training text snippets; ii) through experiments, we show that these methods identify responsive text snippets (rationales) better than traditional document-level training classification.

Another contribution includes, using these innovative training methods to build models that we know result in higher precision when compared to classic document-level text trained models [6]. This contribution helps address the burden of having human reviewers identify text snippets for training.

The rest of the paper is organized as follows. Section II discusses previous work in explainable text classification. We describe the two proposed methods in Section III and report the experimental results in Section IV. We summarize our findings and conclude in Section V.

## II. PREVIOUS WORK

Research in Explainable Machine Learning has focused on two main approaches: model-based explanations and prediction-based explanations. In a model-based approach, since certain types of machine learning models, such as decision tree models or If-Then rule-based models or linear models are inherently easy to interpret (or "explain") using a human's point of view, the ultimate goal is to create machine learning models that are either based on interpretable models or that can be approximated or reduced into interpretable model components. Complex models, such as deep learning models like multilayer neural network models, non-linear SVM, or ensemble models are not directly human understandable and require implementing a more sophisticated approach to interpret the models.

Another explainable machine learning approach is prediction-based and creates an "explanation" for each individual prediction generated by a complex model: i.e., to explain the outcome of the model. Generally, a prediction-based explanation approach provides an explanation as a vector with real value weights, each for an independent variable (feature), indicating the extent to which it contributes to the classification. This approach is not ideal for text classification, due to the high dimensionality in the feature space. In many text classification tasks, a document belongs to a category, most likely because some passages of the text in the document support the classification. Therefore, a small portion of the document text could be used as evidence to justify the classification decision in these text classification tasks.

Recent research found that a prediction-based approach is often used to identify snippets of text as an explanation for the classification of a document. A text snippet that explains the classification of a document is called a 'rationale' for the document in [4][5]. From a machine learning perspective, annotated training text snippets provide more effective labeled input due to their targeted evidence relating to the relevance of the decision. Several research examples show that annotated training text snippets could be used to improve model performance [4]. Zaidan, et al. [4] proposed a machine learning method to use annotated training text snippets in documents to boost text classification performance. In their method, the labeled documents, together with human-annotated text snippets, were used together as training data to build a text classification model using SVM. The results demonstrated that classification performance significantly improved with annotated text snippets over the baseline SVM variants using a document's entire text. In [6's] experiments, classification models trained using only annotated text snippets and sampled non-responsive text snippets, performed significantly better than models trained using entire responsive and non-responsive document text when classifying responsive rationales from not responsive rationales. Zhang et. al. [7] used text snippets to augment convolutional neural network models for text classification by boosting the text snippets' contribution to the aggregated document representation. They found that the augmented model consistently outperforms strong baselines.

An essential part of prediction-based explainable text classification is to generate rationales for text classification that serve as the explanation of the prediction. Zaidan et. al. [8] is among the first research to model human annotators to identify contextual rationales in a document. They used a generative approach and trained models using human-annotated text snippets. Lei et al. [9] proposed a neural network approach to generate rationales for text classification. Their approach combined two components, a rationale generator and a rationale encoder, which were trained to operate together. The generator identified a set of candidate rationales and the encoder decided the classification of each candidate rationale. The proposed approach was designed to provide explanations of multi-aspect sentiment analysis and evaluated using manually annotated test cases. The results showed that their approach outperformed the baseline by a significant margin. Chhatwal et. al. [6] used models trained using either entire documents or annotated text snippets to identify rationales within overlapping document snippets. They demonstrated that simple models can successfully identify rationales at close to 50% recall by only reviewing the top two ranked snippets of each document.

In addition to using rationales directly from the target document text to explain the prediction, there are approaches that derive rationales from other sources. In [10], Martens and Provost described a method in which the explanation of a document classification was the minimal set of the most relevant words, such that removing those words from the document would change the classification of the document. In the popular LIME tool, predictions of a black box text classifier are explained by creating an interpretable model that provides explanations in the form of positive and negative class words that are most relevant to the individual predictions [11]. Li et al. [12] proposed a method for finding small pieces of text as the explanation for a document's classification using neural networks. A small piece of text could be a word or a phrase or a sentence. They proposed a reinforcement learning method for finding the piece of text with the minimum number of words that changes the classification after the selected text is removed from the document. Mahoney et al. [13] introduced a framework for explainable legal text classification. In their framework, three approaches were proposed. The first approach simply finds words with large weights as the explanation. The second approach uses the document-level text model to rank all text snippets – top-ranked snippets (or rationales) are considered the document's explanation. In the last approach, text snippets are ranked based on the reduction of the document probability score when the snippet is removed from the document – top-ranked snippets are selected as the explanation. Other related research includes deriving precise attribute (or aspect) value predictions to serve as the explanation of the predictions [14].

III. METHODS FOR IDENTIFYING RESPONSIVE RATIONALES

The main purpose of explainable text classification in our study is to provide additional information (explanations) about a predictive model's document classification decision and to help attorneys more effectively and efficiently identify responsive documents during legal document review . As with many other explainable text classification approaches, we use the prediction-based approach instead of the model-based approach. Additionally, we are only interested in generating explanations for responsive documents, therefore we focus on documents identified as responsive. We assume that we have only labeled documents, but no annotated text snippets.

An explanation of a responsive document is one or more text snippets, referred to as rationale in [5], in the responsive document. Explainable predictive coding sets out to build a method to estimate the following probability:

$$Pr(r=Rationale|x, y=Responsive) \times Pr(y=Responsive |x), \quad (1)$$

where x is a document, y is the model-labeled designation of the document (for example, 'responsive' or 'not responsive') and r is a text snippet from x.

A simple method for finding rationales is using a document-level text classification model to identify rationales. Authors in [6] compared the performance of document-level text classification models and snippet-level text classification models trained using labeled snippets. Document-level models did not perform as well as the snippet-level models, but their results were not too much worse than the snippet-level models. In this paper, we use document-level text models as our baseline models.

Training documents in a legal document review matter generally can contain tens of thousands of tokens (words) and it is likely that most of the tokens in these long documents do not contain responsive content. Therefore, document-level models trained using such documents may be less accurate in identifying short responsive text snippets than a method that derives training data using snippet-level text.

Throughout the rest of this section, we describe our two proposed methods for more accurate rationale detection. The first method applies a document-level text model to score all overlapping text snippets of all training documents. A text snippet is a sequence of N words from the document and two consecutive snippets are overlapped with N/2 words. Then, a set of high scoring snippets from responsive training documents are selected as responsive training snippets and a set of randomly selected snippets from non-responsive documents are used as non-responsive training snippets. A snippet-level detection model is then trained using these training snippets. Algorithm 1 describes the algorithm. SelectResp is responsive snippet selection algorithm. This method is referred as to Snippet Model Method throughout the rest of the paper.

**Algorithm 1: Snippet Model Method Algorithm**
1. **Input**: RespDocs, NonrespDocs
2. **Output**: Snippet Detection Model
3. Model = Train(RespDocs, NonrespDocs)
4. Let N be the snippet size
5. ScoredSnpts = []
6. For d ∈ RespDocs
7.     Snpts = GetAllSnpts(N, d)
8.     ScoredSnpts = ScoredSnpts + Model.Score(Snpts)
9. RespSelected = SelectResp(ScoredSnpts)
10. For d ∈ NonrespDocs
11.     Snpts = GetAllSnpts(N, d)
12.     nonrespSnpts = nonrespSnpts + Snpts
13. NonrespSelected = RandSelect(nonRespSnpts)
14. Model = Train(RespSelected, NonrespSelected)
15. Return Model

The selection of responsive training snippets from responsive documents is not a trivial task. First, we aim to select only snippets that are very likely responsive and contain high probability scores. Second, we aim to cover as many responsive documents as possible, namely we select responsive snippets from as many responsive documents as possible so that we can target a broad range of responsive training content. However, many responsive documents may not include snippets with high probability scores. Therefore, the selected responsive training snippets may only cover a subset of all responsive training documents. Algorithm 2 summarizes the algorithm for selecting responsive training snippets. In our experiments, minScoreTh = 0.8 and maxNum = 500. In the future, we intend to develop a more sophisticated responsive training snippet selection algorithm.

**Algorithm 2: Responsive Snippet Selection Algorithm**
1. **Input**: ScoredSnpts
2. **Ouput**: SelectedSnpts
3. Descending_Score_Sort(ScoredSnpts)
4. SelectedSnpts = []
5. SelectedDocs = []
6. nSelectedSnpts = 0
7. For s ∈ *ScoredSnpts*
8.     If $s.score \geq 0.5$ and s.doc not in SelectedDocs
9.         SelectedSnpts = SelectedSnpts + s
10.        SelectedDocs = SelectedDocs + s.doc
11.        nSelectedSnpts = nSelectedSnpts + 1
12. For s ∈ *ScoredSnippets*
13.    If s Not In SelectedSnpts and s.score ≥
14.       minScoreTh and nSelectedSppts ≤ maxNum
15.          SelectedSnpts = SelectedSnpts + s
16.          nSelectedSnpts = nSelectedSnpts + 1
17. Return SelectedSnpts

**Algorithm 3: Iterative Snippet Method Algorithm**
1. **Input**: RespDocs and NonRespDocs
2. **Output**: Iterative Snippet Detection Model
3. Model = Train(RespDocs, NonrespDocs)
4. Let N be the snippet size
5. Finished = False
6. Repeat
7.     ScoredSnpts = []
8.     For d ∈ RespDocs
9.         Snpts = GetAllSnpts(N, d)
10.        ScoredSnpts = ScoredSnpts +
11.           Model.Score(Snpts)
12.    RespSelected = SelectResp(ScoredSnpts)
13.    For d ∈ NonrespDocs
14.        Snpts = GetAllSnpts(N, d)
15.        nonrespSnpts = nonrespSnpts + Snpts
16.    NonrespSelected = RandSelect(nonRespSnpts)
17.    Model = Train(ResponpSelected, NonrespSelected)
18.    If N > MinSnippetSize
19.        $N = \frac{N}{2}$
20.    If N < MinSnippetSize
21.        N = MinSnippetSize
22.    Else Finished = True
23. Until Finished
24. Return Model

The second method uses an iterative algorithm approach, which iteratively refines the number of tokens in the training text snippets. It starts with the document-level model and applies it to identify large responsive training snippets – e.g., 1,000 words. During each iteration, a new snippet-level model is trained using a new iteration of training snippets. Then that iteration's snippet-level model is applied to identify new set of training snippets but where the token size of the snippet is reduced by half. This process repeats until the snippet size meets the user-defined minimal size. This approach is referred as to Iterative Snippet Model Method. Algorithm 3 summarizes the algorithm for the iterative snippet method.

IV. EXPERIMENTS

In this section, we describe our experiments and report our experimental results on three datasets from three real legal document review matters. We describe the three datasets and the design of the experiments in Sections IV.A and IV.B, respectively. The results are reported in Section IV.C.

*A. The Dataset*

*a)* The data for the experiments was randomly selected from attorney-labeled responsive and nonresponsive documents from the three legal document review matters. The randomly selected documents from each legal document review matter dataset were randomly divided into training and testing sets. Each matter's dataset included emails, Microsoft Office documents, PDFs, and other text-based documents. Table 1 summarizes the number of responsive and nonresponsive documents in the training and testing sets for each of the three datasets. As mentioned in the introduction, we do not have labeled text snippets – the labels were applied at a document level.

TABLE I. DATASET STATISTICS

| Dataset | Training Documents | | Testing Documents | |
|---|---|---|---|---|
| | *Resp* | *Nonresp* | *Resp* | *Nonresp* |
| A | 2,000 | 10,000 | 8,000 | 40,000 |
| B | 4,000 | 8,000 | 12,000 | 24,000 |
| C | 2,000 | 6,000 | 6,300 | 21,000 |

*B. Experiment Design*

The purpose of these experiments was to empirically compare the two snippet model methods (the Snippet Detection Model and the Iterative Snippet Detection Model) with the baseline method ( the Document-Level Model). The minimum snippet size was 50 tokens for all experiments. The final snippet detection models were trained and tested using snippets that contained 50 tokens (words). The Iterative Snippet Model Method began with 1,000 token snippets and each new iteration reduce the snippet token size by half until the snippet size was set to 50 tokens.

All models were trained using the logistic regression algorithm. Prior studies demonstrated that Logistic Regression performs very well on legal document review matters [15, 16]. The bag of words method with 3-grams was used to represent the documents and text snippets. We applied Information Gain to select 2,000 tokens as features with normalized token frequency as feature values.

Since we do not have labeled text snippets, we cannot evaluate these models using conventional performance metrics such as precision and recall. Instead, we evaluated these models by measuring the reduction in the document's score when their identified rationales (the resulting text snippets) were removed and the number of responsive documents with detected rationales.

We tested our models using the population of responsive documents in the testing set that received a document-level model score of greater than or equal to .5. .5 is the score that Logistic Regression uses to divide the two classes. In legal document review, .5 may not always be the best score to divide the two classes but without having a typical set of test data, we decided this was a useful metric to divide the snippets between an accurate explanation and a mistake.

We assume only documents with scores greater than or equal to 0.5 would be detected as responsive documents by the document-level model and that we only need to generate rationales for the documents the model would have detected as responsive documents. For this set of test documents, we segmented the documents into 50 token overlapping snippets and the models were then applied to each snippet to generate a probability score. High scoring text snippets in the documents were the snippets that we labeled as the rationales for the documents.

Next, the rationales were removed from the document's text, and we rescored the documents without their rationale/s using the Document-Level Model. The score reduction for each document was computed as the difference between the scores of the original document and the document with its rationales removed. The model with the higher average score reduction is considered the better model for identifying rationales.

*C. Results*

Tables 2, 3, and 4 summarize the average score reduction for each of the three methods for each of the three datasets A, B, and C, respectively. Each Snippet Score Threshold (Snippet Score TH) establishes a set of responsive documents. The set of documents defined by the threshold *th* consists of the responsive documents whose document score is greater than or equal to 0.5 and their largest snippet scores are in the range [*th*, *th*+0.1) except for *th* = 0.9. For *th* = 0.9, the range is [0.9, 1]. *#Doc* is the number of responsive documents in the corresponding document set established by the threshold. *Avg Doc Score* is the average document score of all responsive documents included in the threshold set. For each responsive test document with a document-level model score of greater than or equal to .5, we identified its largest snippet score in the threshold range [*th*, *th*+0.1) or [0.9, 1] and we removed all snippets with scores in the range and then applied the Document-Level model to assign probability scores to these documents. *Avg Doc Score with Snippet Rmd* is the average document score of the documents within the threshold set with their snippets removed. *Avg Doc Score Reduction* is the difference between *Avg Doc Score* and *Avg Doc Score with Snippet Rmd* and is the average score reduction for the Snippet Score TH. In row *Tot/Avg*, #Docs is the total number of responsive documents whose document score is greater than or equal to 0.5 and have snippet scores greater than or equal to 0.5., *Avg Doc Score* is the average score of all documents with a snippet score greater than or equal to 0.5., and *Avg Doc Score with Snippet Removed* is the average score of all documents with their snippets removed.

*1) Number of Responsive Documents:* Across the three datasets, we can see both the Snippet Models and Iterative Snippet Models identified 50% more responsive documents (#Doc) than the Document-Level models at the [0.9, 1] score threshold. In Datasets A and C, the Snippet and Iterative Snippet models identified more total responsive documents than the Document-Level model with snippet scores from threshold 0.5 to 1, while the Document-Level model identified more total responsive documents in the same threshold for Dataset B. The results clearly show that the Snippet and Iterative Snippet models performed significantly better than the Document-Level models when identifying rationales (high scoring snippets). The Snippet models generated slightly more responsive documents than the Iterative Snippet models with snippet scores in the [0.9, 1] threshold for Datasets A and B, but slightly fewer responsive documents in Dataset C.

TABLE II. DATASET A – SCORE REDUCTION STATISTICS

| Snippet Score TH | Document-Level Model | | | | Snippet Model | | | | Iterative Snippet Model | | | |
|---|---|---|---|---|---|---|---|---|---|---|---|---|
| | #Doc | Avg Doc Score | Avg Doc Score With Snippet Rmd | Avg Doc Score Reduction | #Doc | Avg Doc Score | Avg Doc Score With Snippet Rmd | Avg Doc Score Reduction | #Doc | Avg Doc Score | Avg Doc Score With Snippet Rmd | Avg Doc Score Reduction |
| [0.9, 1] | 2,140 | 0.96 | 0.46 | 0.5 | 3,598 | 0.91 | 0.17 | 0.74 | 3732 | 0.91 | 0.21 | 0.7 |
| [0.8, 0.9) | 419 | 0.89 | 0.44 | 0.45 | 165 | 0.78 | 0.39 | 0.39 | 170 | 0.79 | 0.44 | 0.35 |
| [0.7, 0.8) | 273 | 0.84 | 0.44 | 0.4 | 114 | 0.71 | 0.37 | 0.34 | 67 | 0.77 | 0.41 | 0.36 |
| [0.6, 0.7) | 227 | 0.85 | 0.41 | 0.43 | 62 | 0.72 | 0.32 | 0.4 | 45 | 0.79 | 0.46 | 0.33 |
| [0.5, 0.6) | 186 | 0.84 | 0.38 | 0.46 | 45 | 0.72 | 0.43 | 0.29 | 44 | 0.74 | 0.37 | 0.37 |
| Tot/Avg | 3,245 | 0.92 | 0.45 | 0.47 | 3,984 | 0.89 | 0.19 | 0.7 | 4,056 | 0.9 | 0.23 | 0.67 |

TABLE III. DATASET B – SCORE REDUCTION STATISTICS

| Snippet Score TH | Document-Level Model | | | | Snippet Model | | | | Iterative Snippet Model | | | |
|---|---|---|---|---|---|---|---|---|---|---|---|---|
| | #Doc | Avg Doc Score | Avg Doc Score with Snippet Rmd | Avg Doc Score Reduction | #Doc | Avg Doc Score | Avg Doc Score with Snippet Rmd | Avg Doc Score Reduction | #Doc | Avg Doc Score | Avg Doc Score with Snippet Rmd | Avg Doc Score Reduction |
| [0.9, 1] | 936 | 0.92 | 0.62 | 0.3 | 1,504 | 0.87 | 0.51 | 0.36 | 1,611 | 0.86 | 0.51 | 0.35 |
| [0.8, 0.9) | 637 | 0.83 | 0.52 | 0.31 | 341 | 0.77 | 0.53 | 0.24 | 239 | 0.79 | 0.55 | 0.24 |
| [0.7, 0.8) | 610 | 0.77 | 0.44 | 0.33 | 219 | 0.75 | 0.54 | 0.21 | 165 | 0.77 | 0.50 | 0.27 |
| [0.6, 0.7) | 490 | 0.71 | 0.41 | 0.3 | 170 | 0.74 | 0.5 | 0.24 | 198 | 0.78 | 0.6 | 0.18 |
| [0.5, 0.6) | 327 | 0.68 | 0.39 | 0.29 | 188 | 0.75 | 0.54 | 0.21 | 127 | 0.76 | 0.57 | 0.19 |
| Tot/Avg | 3,000 | 0.81 | 0.50 | 0.31 | 2,422 | 0.83 | 0.52 | 0.31 | 2,340 | 0.84 | 0.52 | 0.32 |

TABLE IV. DATASET C – SCORE REDUCTION STATISTICS

| Snippet Score TH | Document-Level Model | | | | Snippet Model | | | | Iterative Snippet Model | | | |
|---|---|---|---|---|---|---|---|---|---|---|---|---|
| | #Doc | Avg Doc Score | Avg Doc Score with Snippet Rmd | Avg Doc Score Reduction | #Doc | Avg Doc Score | Avg Doc Score with Snippet Rmd | Avg Doc Score Reduction | #Doc | Avg Doc Score | Avg Doc Score with Snippet Rmd | Avg Doc Score Reduction |
| [0.9, 1] | 1,311 | 0.93 | 0.57 | 0.36 | 3,101 | 0.84 | 0.28 | 0.56 | 3,047 | 0.84 | 0.36 | 0.48 |
| [0.8, 0.9) | 705 | 0.83 | 0.47 | 0.36 | 210 | 0.70 | 0.44 | 0.26 | 160 | 0.72 | 0.46 | 0.26 |
| [0.7, 0.8) | 470 | 0.76 | 0.44 | 0.32 | 80 | 0.69 | 0.45 | 0.24 | 83 | 0.69 | 0.44 | 0.25 |
| [0.6, 0.7) | 403 | 0.73 | 0.43 | 0.3 | 56 | 0.66 | 0.43 | 0.23 | 54 | 0.71 | 0.49 | 0.22 |
| [0.5, 0.6) | 312 | 0.68 | 0.38 | 0.3 | 23 | 0.64 | 0.41 | 0.23 | 48 | 0.67 | 0.41 | 0.26 |
| Tot/Avg | 3,201 | 0.83 | 0.49 | 0.34 | 3,470 | 0.82 | 0.30 | 0.52 | 3,392 | 0.83 | 0.37 | 0.46 |

*2) Average Score Reduction:* For Dataset A and C, the Snippet and Iterative Snippet models achieved much higher average score reductions than the document-level models when rationales were removed from the documents. For these datasets, the average score reductions for the Snippet and Iterative Snippet models was 0.7 and 0.67, respectively, while the average score reduction was 0.47 for the Document-Level model. For Dataset C, the average score reductions for the Snippet and the Iterative Snippet models were 0.52 and 0.46, respectively, while the average score reduction was 0.34 for the Document-Level model. For Dataset B, all three models achieved similar average score reductions – slightly higher than 0.3. Lastly, for Datasets A and C, the Snippet models achieved slightly higher average score reductions than the Iterative Snippet models.

*3) Average Doc Score with Snippet Removed:* In most cases, for the Snippet and the Iterative Snippet models, the removal of snippets with higher snippet scores resulted in larger average score reductions. The average score reductions in the [0.9, 1] threshold for the Snippet and Iterative Snippet models are much higher than the average score reductions in other threshold ranges.

Figure 1 shows the precision and recall curves for the Document-Level model when it is applied to the three datasets to make a document classification decision of responsive or not responsive. The performance of the model on Dataset A was better than on Dataset C. For Dataset B, the model's performance was significantly worse than on Datasets A and C. Similarly, when applying snippet detection, the three methods performed best on Dataset A and they performed better on Dataset A and C than they did on Dataset B.

Both Snippet and Iterative Snippet models identified responsive documents significantly better than the Document-Level model on Dataset A and C. The models' performance on Dataset B was significantly worse when compared to Datasets A and C. We observed that the Document-Level model's performance on the document classification task has an important impact on the rationale detection performance for the two proposed snippet model methods. When the document model is accurate, the snippet models performed much better than the Document-Level model in identifying rationales. Responsive snippets for training a snippet model are identified using the document-level model, therefore training snippets would include many misidentified responsive snippets when the document model is not accurate. These incorrectly identified responsive training snippets would degrade the performance of the trained snippet model.

Snippet models almost always performed slightly better than iterative snippet models. This is probably due to the error propagations through the sequence of models developed. An iterative snippet model starts from a document model, which is never %100 accurate, Then, a sequence of snippet models is generated by reducing the size of snippets by half. The classification errors propagate through this sequence of models.

Table 5 reports the average number of tokens in a document and the average number of tokens removed as identified rationales. From the table, we can see snippet models always removed more tokens than document models. This means snippet models were able to detect more rationales than document models. It implies that higher document score reductions achieved by snippet models are partly caused by more rationales identified by snippet models.

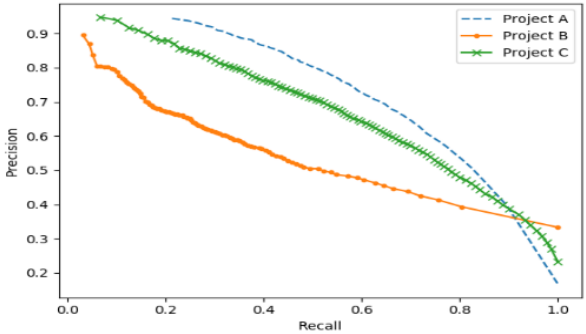

Fig. 1. Precision and Recall Curves for Document-Level Models

TABLE V. AVERAGE NUMBER DOCUMENT TOKENS AND NUMBER OF TOKENS REMOVED

| Project | Document Model | | Snippet Model | | Iterative Snippet Model | |
|---|---|---|---|---|---|---|
| | *Average Number of Tokens in a Document* | *Average Number of Tokens Removed* | *Average Number of Tokens in a Document* | *Average Number of Tokens Removed* | *Average Number of Tokens in a Document* | *Average Number of Tokens Removed* |
| A | 1,020 | 94 | 1,275 | 190 | 898 | 176 |
| B | 538 | 67 | 623 | 97 | 514 | 92 |
| C | 1,304 | 113 | 1,305 | 218 | 1,176 | 151 |

## V. SUMMARY AND CONCLUSIONS

This paper proposed two innovative methods for explainable text classification in legal document review under the assumption that there are not labeled text snippets to train rationale detection models. Specifically, the two methods are the Snippet Model method and Iterative Snippet Model method. We conducted experiments, using three real legal matter datasets, to evaluate the two methods and compared them with the popular Document-Level model method. We evaluated the performance of each method without labeled rationales since creating labeled rationales in the legal domain is not practical due to the time-consuming nature of identifying the rationales. Instead of using conventional performance metrics such as precision and recall, we used the reduction of the document score achieved by removing the models' identified rationales.

The Document-Level Model Method is the simplest to implement because it requires no extra work and is our baseline for assessing performance. The Snippet Model Method is also simple but takes more time to score each snippet. The Iterative Snippet Model takes significantly more time to implement because of the training iterations required to achieve the final snippet model. Each method identified rationales reasonably well. Both snippet methods performed better than the Document-Level method when the document models are accurate. Both snippet methods achieved greater score reductions and identified high-scored snippets for more documents than the Document-Level model. They also identified more rationales for each document. The Snippet Model Method performed slightly better than the Iterative Snippet Model Method.

The results demonstrate that it is feasible to build machine learning models that can automatically identify rationales without using annotated text snippets for training. This is an exciting result given how time-consuming generating labeled training text snippets is for lawyers and that a model trained with text snippets outperforms one that is trained using an entire document's text. Incremental improvement in precision at certain recall rates can have a significant impact on the cost of the document review process. Consider a legal review matter that used a text classification model to identify 1 million responsive documents for review. If the precision of that model could be improved by 5 percent, it could result in a cost savings of at least $50,000.

We plan to conduct more experiments using additional datasets and further, we intend to explore more advanced machine learning technologies to continue evolving our understanding of categorizing rationales for training models and explaining classification results. In addition to identifying fixed length text snippets as rationales, we plan to explore segmenting a document into sentences, groups of sentences, and paragraphs and then use this segmenting to identify rationales as a sentence, a group of sentences, or a paragraph. We also intend to develop more advanced methods for selecting responsive and nonresponsive training snippets. Human labeled documents can almost always improve performances of trained models. We shall conduct experiments with human reviewers involved in the loop. The document model will provide a small list of high score snippets and human reviewers labels all snippets in the list. Then, labeled snippets are used to train a snippet model.